# Physical meaning of the Ewald sum method

T. R. S. Prasanna

Department of Metallurgical Engineering and Materials Science

Indian Institute of Technology, Bombay

Mumbai – 400076

India

The electrostatic potential and energy of point charges in a real crystal, in the presence of thermal vibrations, is obtained as a special case of the Fourier method. Incorporating the role of thermal vibrations in electrostatic energy calculations leads to the physical meaning of the Ewald sum method. The Ewald summation method determines the electrostatic potential and energy of point charges in a crystal at a temperature that is obtained from the width of the Gaussian and not at 0 K. For values of the width of the Gaussian commonly recommended for computational convenience temperatures exceed 10000 K.

## 1. Introduction

Determining the electrostatic potential and energy of an idealized crystal that contains positive and negative point charges is a classical problem that has been addressed for almost a century starting with Madelung (1918) and Ewald (1921) [1,2]. The Ewald summation method is also the standard method used to evaluate the repulsive inter-nuclear electrostatic energy, $E_{n-n}$, in electronic structure calculations. There are two equivalent derivations of the Ewald method, both of which are described in several textbooks [3-9], one of which is more transparent. Briefly, the latter method [3-7] consists of smearing the nuclear point charge into a Gaussian function in the first step and calculating the resulting potential by also accounting for the self-potential. In the second step, the potential due to a negative Gaussian (that cancels the positive Gaussian of the first step) and a positive point charge is added. The net potential due to both the steps gives the electrostatic potential at any lattice point due to all other point charges in the crystal. The final expression requires computation in both real and Fourier space and the width of the Gaussian is selected *purely for computational convenience*.

In 1952, Bertaut [10] pointed out that a basic result from electrostatics, viz. the potential due to non-overlapping spherically symmetric charge distribution is identical to that of point charges, can be used to determine the electrostatic potential and energy due to nuclear point charges. There was some controversy [11,12] regarding the derivation of the relevant expressions. However, this controversy has been resolved [13] as it has been shown that both expressions are valid for the particular problem at hand, even though the rates of convergence may be different. These methods require the evaluation of only one series in Fourier space and hence will be referred to as Fourier methods. In these methods as well, the choice of the spherically symmetric function into which the nuclear point charge is expanded is based solely on computational convenience.

However, in all real crystals, lattice vibrations are present and have a significant effect on various physical properties. The effect of thermal vibrations on the nuclear charges and the electron distributions is well established in diffraction theory. In the harmonic approximation, with the assumption of rigid ions or pseudo atoms, the effect of thermal vibrations is to change the Fourier Transform of the charge density, the atomic scattering factor, $f_j$ to $f_j e^{-M_j}$ [14-18] where $M_j$ is the Debye-Waller factor (DWF). Thus the main effect of thermal vibrations can be described as "*To an extremely good approximation, the scattering averaged over the instantaneous distributions is equivalent to the scattering of the time-averaged distribution of the scattering matter*"[14]. Similar arguments are applicable in analysis of electron diffraction data [17,18] where the effect of thermal vibrations on both the electron and nuclear charge distributions are considered. The Fourier Transform of nuclear charge is altered from $Z_j$ to $Z_j e^{-M_j}$ due to thermal vibrations.

We have recently shown that diffraction data impose severe constraints on theoretical models for order-disorder phase transitions in alloys [19]. It is essential to incorporate the role of thermal vibrations for a correct understanding of order-disorder phase transitions. All existing models must be modified to explicitly include a temperature dependent interaction parameter before their predictions can be compared with diffraction data. The simplest mean-field Bragg-Williams model was modified to make the ordering energy temperature dependent so as to enable its predictions to be compared with diffraction data. In order to obtain a simple form for the temperature dependence of the ordering energy for the simplest Modified Bragg-Williams model, we had in passing identified the Gaussian in the Ewald sum method with thermal vibrations [19].

In all studies on electrostatic potential and energy of crystals till date, including recent studies [20-22], whether based on Ewald method or Fourier method, the emphasis has been on computational issues. No attempt has been made to incorporate the role of thermal vibrations that are universally present in real crystals. In this paper, we consider the consequences of incorporating the role of two physical features that are universal to all real crystals on the analysis of electrostatic potential and energy. These physical features are 1) the presence of thermal vibrations and 2) the first phase transition temperature, $T_p$, which could be either the melting point or the solid-solid transition temperature. While the derivations are simple, the implications of the results are very significant for the physical meaning of the Ewald method.

**2. Electrostatic potential and energy of crystals in the presence of thermal vibrations**

The role of thermal vibrations is to smear the nuclear point charge into a time-averaged continuous distribution. *Therefore, there are no point charges but only continuous charge distributions in a real crystal.* This is true even at 0 K due to the presence of zero-point vibrations. *Hence, the problem to be addressed is the electrostatic potential and energy of* *charge distributions and not point charges.* This is a simple problem and it solution can be easily obtained as shown below.

We begin with the standard result [17,18] from diffraction theory that the Fourier Transform of the nuclear point charge in the presence of thermal vibrations is given by

$$\sigma_G = Z_j \, e^{-M_j} \quad (1)$$

where $Z_j$ is the charge on nucleus $j$ and $M_j$ is the DWF. As is well known [16], the DWF can be represented in different ways as

$$M = 8\pi^2 \langle u_s^2 \rangle \left(\frac{\sin\theta}{\lambda}\right)^2 = B \left(\frac{\sin\theta}{\lambda}\right)^2 \quad (2)$$

where $\langle u_s^2 \rangle$ is the mean-squared displacement and B is the isotropic temperature factor. At high temperatures (T ≥ $\Theta_D$), the DWF is also given by [4,16]

$$M = \frac{6h^2 T}{mk_B \Theta_D^2} \left(\frac{\sin\theta}{\lambda}\right)^2 \quad (3)$$

We note that $(\sin\theta/\lambda)^2 = \boldsymbol{G}^2/4$ as is usually used in diffraction theory and follows from the relation between the real space and reciprocal space vectors, $\boldsymbol{a_i} \cdot \boldsymbol{a_j^*} = \delta_{ij}$. The Fourier Transform of the nuclear point charge in the presence of thermal vibrations, Eq.1, is now given by [17]

$$\sigma_G = Z_j\, e^{-2\pi^2 \langle u_s^2 \rangle G^2} = Z_j\, e^{-\frac{BG^2}{4}} \qquad (4)$$

Assuming the mean-square displacement, $\langle u_s^2 \rangle$, to be isotropic, as is the first approximation in diffraction theory, and taking the inverse Fourier Transform of Eq.4 [6], *the nuclear point charge is transformed into a spherically symmetric Gaussian charge distribution due to thermal vibrations at finite temperatures* and is given by

$$\sigma(r) = \frac{Z_j}{((2\langle u_s^2 \rangle)^{3/2} \pi^{3/2})}\, e^{-\frac{r^2}{2\langle u_s^2 \rangle}} \qquad (5)$$

Thus, for real crystals, Eq.5 must be used to determine the electrostatic potential and energy of crystals at finite temperatures. The existing Fourier method formalism [13] that is used for point charges must be modified for continuous charge distributions.

From standard texts [17] in electron microscopy, the Fourier Transform of the nuclear charge density in the unit cell, or the structure factor, is given by

$$F_G = \sum_{j=1}^{N} Z_j\, e^{2\pi i G \cdot r_j} \qquad (6)$$

In the presence of thermal vibrations and assuming the same value of M or $\langle u_s^2 \rangle$ for all species, the Fourier Transform of the nuclear charge density at finite temperatures is given by [17,18]

$$F_G(T) = \sum_{j=1}^{N} Z_j\, e^{-\frac{BG^2}{4}}\, e^{2\pi i G.r_j} = F_G\, e^{-\frac{BG^2}{4}} \tag{7}$$

The charge density in the unit cell at finite temperatures is given by the inverse Fourier Transform as

$$\rho(r,T) = \frac{1}{v} \sum_{G \neq 0} F_G\, e^{-\frac{BG^2}{4}}\, e^{-2\pi i G.r} \tag{8}$$

Using the Poisson Equation, the Fourier Transform of the total potential at finite temperatures is given by

$$V_G^{tot}(T) = \frac{1}{\pi v}\, \frac{F_G\, e^{-\frac{BG^2}{4}}}{G^2} \tag{9}$$

The total potential in the unit cell at finite temperatures is given by

$$V^{tot}(r,T) = \frac{1}{\pi v} \sum_{G \neq 0} \frac{F_G\, e^{-\frac{BG^2}{4}}}{G^2}\, e^{-2\pi i G.r} \tag{10}$$

Eq.10 is the core potential due to nuclear charges that is used in diffraction theory [17,18]. To calculate the electrostatic energy, we must find the potential at the site of one ion due to all other ions for which we must subtract the contribution from the self-potential from the ion at that site. The Fourier Transform of the self-potential of ion $j$ is given by [13]

$$V_G^S(T) = \frac{1}{\pi} \frac{Z_j \, e^{-\frac{Bg^2}{4}}}{g^2} \tag{11}$$

The self-potential of ion $j$ at finite temperatures at any point $r$ is given by

$$V_s(r,T) = \frac{Z_j}{\pi} \int \frac{e^{-\frac{Bg^2}{4}}}{g^2} \, e^{-2\pi i g \cdot r} \, d^3g \tag{12}$$

The self-potential at the site of any ion is found by setting $r = 0$ is Eq.12 and is obtained as

$$V_s(0,T) = Z_j \frac{2}{((2\langle u_s^2 \rangle)^{1/2} \pi^{1/2})} \tag{13}$$

The potential at the site of any ion due to all other ions is given by

$$V_{int}(\mathbf{r_j}) = \frac{1}{\pi v} \sum_{G \neq 0} \frac{F_G \, e^{-\frac{BG^2}{4}}}{G^2} \, e^{-2\pi i G \cdot r_j} - Z_j \frac{2}{((2\langle u_s^2 \rangle)^{1/2} \pi^{1/2})} \tag{14}$$

The total electrostatic energy at finite temperatures is given by $\frac{1}{2} \int \rho(r,T) \, V^{tot}(r,T) \, d^3r$ and in terms of the Fourier components can be written as $\sum_{G \neq 0} \rho_G(T) \, V_G(T)$

$$E^{tot}(T) = \frac{1}{2\pi v} \sum_{G \neq 0} \frac{|F_G|^2 \, e^{-\frac{2BG^2}{4}}}{G^2} \tag{15}$$

Eq.16 contains the self energy of all ions as well, which must be subtracted. The self energy of each ion $j$ is given by

$$E^s(T) = \frac{Z_j^2}{2\pi} \int \frac{e^{-\frac{2Bg^2}{4}}}{g^2} d^3g = \frac{Z_j^2}{\sqrt{2\pi}} \frac{1}{(2\langle u_s^2 \rangle)^{1/2}} \quad (16)$$

Therefore, the electrostatic energy obtained by incorporating the role of thermal vibrations that are present in a real lattice is given by

$$E^{int}(T) = \frac{1}{2\pi v} \sum_{G \neq 0} \frac{|F_G|^2 e^{-\frac{2BG^2}{4}}}{G^2} - \frac{1}{\sqrt{2\pi}(2\langle u_s^2 \rangle)^{1/2}} \left( \sum_j Z_j^2 \right) \quad (17)$$

In all the above equations, we have preserved the quantity $(2\langle u_s^2 \rangle)$ because it equals $\eta^2$, where $\eta$ is the width of the Gaussian in Tosi [6] and Brown [7]. This allows the width of the Gaussian function in standard derivations to be related easily to the mean-squared displacement of ions.

We note that our derivation till Eq.14, which gives the potential at the site of any ion due to all other ions, is identical to the derivation in Ref.13. In particular, Eq.14 above is exactly equal to Eq.17 of Ref.13 with $\phi(h)$ in the latter is replaced by $e^{-\frac{BG^2}{4}}$ and also changing the reciprocal lattice variable.

The expression for the electrostatic energy obtained from Eq.20 of Ref.13 is obtained as

$$E^{int}(T) = \frac{1}{2\pi v} \sum_{G \neq 0} \frac{|\rho_G|^2 e^{-\frac{BG^2}{4}}}{G^2} - \frac{1}{\sqrt{\pi}\,(2\langle u_s^2 \rangle)^{1/2}} \left( \sum_j Z_j^2 \right) \quad (18)$$

where the first term differs in the convergence factor $e^{-\frac{BG^2}{4}}$ from the corresponding first term $e^{-\frac{2BG^2}{4}}$ in Eq.17 and the second term differs from the corresponding term in Eq.17 by a factor of $\sqrt{2}$. Eq.18 has been obtained by assuming a point charge at the ion site, while Eq.17 has been obtained from our derivation where there are only continuous charge distributions and no point charges. Eq.17 above is exactly equal to Eq.1.2 of Ref.12 with $\phi(\mathbf{h})$ in the latter is replaced by $e^{-\frac{BG^2}{4}}$ and also changing the reciprocal lattice variable. Both expressions, Eq.17 and Eq.18, are correct [13] and the reason will be discussed below shortly.

We discuss the actual values of parameters for two systems, Al (metal) and KCl (ionic compound). From data published in literature, the isotropic temperature factor for Al [23], $B_{Al}$ = 2.08 Å$^2$ at 560 K and for KCl [24], $B_K$ = 2.17 Å$^2$ and $B_{Cl}$ = 2.17 Å$^2$ at 295 K. From Eq.2, this leads to mean and root mean-squared (rms) displacement values of $\langle u_s^2 \rangle_{Al}$ = 0.02634 Å$^2$ and $\langle u_s^2 \rangle_{Al}^{1/2}$ = 0.16 Å in Al at 560 K. In KCl, the corresponding values at 295 K are $\langle u_s^2 \rangle_K = \langle u_s^2 \rangle_{Cl}$ = 0.02748 Å$^2$ and $\langle u_s^2 \rangle_K^{1/2} = \langle u_s^2 \rangle_{Cl}^{1/2}$ = 0.165 Å. Given the lattice parameters of Al (fcc), $a_{Al}$ = 4.07 Å [27] and of KCl (fcc), $a_{KCl}$ = 6.29 Å, one-half the nearest neighbor bond length in Al is 1.44 Å and in KCl is 1.57 Å. Since the rms displacements are 0.16 Å, the charge densities will not overlap in either Al or KCl. From Eq.2 and Eq.3, we see that $\langle u_s^2 \rangle$ varies linearly with temperature, $T$, above the Debye temperature. Using Eq.2 and Eq.3, the rms displacement in Al near it melting point (933 K) is $\langle u_s^2 \rangle_{Al}^{1/2}(933\,K) = 0.21$ Å. Similarly, the rms displacements in KCl near its melting point (1043 K) is $\langle u_s^2 \rangle_K^{1/2}$ (1043 K) = $\langle u_s^2 \rangle_{Cl}^{1/2}$ (1043 K) = 0.31 Å. Thus, even near the melting points, the nuclear charge densities will not overlap. We have used Al and KCl only as representative examples. *We can conclude that the nuclear charge densities will not overlap even near the melting points of real crystals.*

An important result from electrostatics [13] is that "*the interaction energy of an array of point charges is the same as for an array of non-overlapping spherically symmetric equivalent charge distributions*". Therefore, Eq.17 and Eq.18, both represent the interaction energy of an array of point charges in a static lattice for all values of the isotropic temperature factor, B, below the melting point.

Therefore, the Bertaut expression [12] for the potential at the site of any ion that is obtained by equating the energy expression for continuous charges to that for point charges, $E^{int} = \frac{1}{2} \sum_j Z_j V_j(r_j)$, can be used to derive another expression for the potential at the site of any ion as

$$V_{int}(r_j) = \frac{1}{\pi v} \sum_{G \neq 0} \frac{\rho_G\, e^{-\frac{2BG^2}{4}}}{G^2} e^{-2\pi i G \cdot r_j} - Z_j \frac{\sqrt{2}}{((2\langle u_s^2 \rangle)^{1/2} \pi^{1/2})} \qquad (19)$$

We now have two expressions, Eq.14 and Eq.19, for the potential at the site of any ion. For NaCl or KCl structure, the Madelung constant, $\alpha_M$, is obtained as $\alpha_M = -\frac{a}{2e} V^{int}(Na^+)$. Since, $\alpha_M$ = 1.747565, the correct value of $V^{int}(Na^+) = -3.49512\ e/a$. Since the results are expressed in units of *e/a*, the correct value of the potential at the site of K$^+$ ion, $V^{int}(K^+) = -3.49512$ *e/a* as well.

We evaluate $V^{int}(K^+)$ using both expressions, Eq.14 and Eq.19, with $\eta^2 = (2\langle u_s^2 \rangle) = a^2/225$ that corresponds to η = *a*/15, which would be the case near the melting point, T$_p$, of KCl. Table 1 summarizes the results of the said calculations. The potential from Eq.19 has converged to the second decimal place by going as far out as |**G**|$^2$ * a$^2$ = 75 while the potential from Eq.14 has not converged to the first decimal place. The superior convergence of Eq.19 is to be expected as it has a faster decaying factor that is the square of the decaying factor in Eq.14. An important difference is that using the Gaussian obtained from thermal vibrations ensures that the decaying factor is always positive or the series converges monotonically. If other expressions for the nuclear charge density are used, often the decaying factor contains trigonometric terms due to which it can take positive or negative values and the series does not converge monotonically.

**3. Physical meaning of the Ewald summation method**

We next discuss the implications of the above results for the Ewald summation method [3-9]. In this method, the point charge is replaced by a Gaussian function, but until now it was only for mathematical convenience. The width of the Gaussian in the Ewald method is selected so that both the series, one in real space and one in reciprocal space, converge quickly.

The first step of the Ewald sum method spreads the nuclear point charge into a Gaussian distribution [3-7]. From diffraction theory [17], and from Eq.1-Eq.5, it is clear that expanding the nuclear point charge into a Gaussian distribution is equivalent to heating the crystal to finite temperatures and incorporating the role of thermal vibrations. Comparing the width of the Gaussian, η, in Ref.6 and Ref.7 with Eq.5 above shows that $\eta^2 = (2\langle u_s^2\rangle)$. It implies that the crystal is at a finite temperature, $T_{Gau}$, that can be determined using Eq.2, Eq.3 and Eq.5. The root mean-squared (rms) displacement values of $\langle u_s^2\rangle_{Al} = 0.02634$ Å² and $\langle u_s^2\rangle_{Al}^{1/2} = 0.16$ Å in Al at 560 K and in KCl, the corresponding values at 295 K are $\langle u_s^2\rangle_K = \langle u_s^2\rangle_{Cl} = 0.02748$ Å² and $\langle u_s^2\rangle_K^{1/2} = \langle u_s^2\rangle_{Cl}^{1/2} = 0.165$ Å as discussed in the previous section. The value usually recommended [6] in the Ewald method to ensure rapid convergence is that the parameter, η, be equal to one-half of the nearest neighbor bond length. Thus, for Al, $\eta_{Al} = 1.44$ Å and $\eta_{KCl} = 1.57$ Å would represent commonly used values of the parameter. We can now calculate the equivalent temperature corresponding to these values. Using Eq.2 and Eq.3 (assuming that $\langle u_s^2\rangle$ varies linearly with temperature) the recommended values [6,7] of the parameter in the Ewald sum method correspond to a temperature $T_{Gau}(Al) = 22000$ K and $T_{Gau}(KCl) = 13000$ K.

The second step involves the superimposition of a negative Gaussian and a point charge *on the result of the first step*, i.e. at $T_{Gau}$. The purpose of the negative Gaussian is to cancel the positive Gaussian of the first step. This would leave the crystal with no charge which is clearly unrealistic. Therefore, the second step of the Ewald sum method is a purely mathematical device to cancel the effects of thermal vibrations and restore point charges in a crystal that is <u>already at $T_{Gau}$</u>. The net result of both the steps is an array of point charges at $T_{Gau}$ and not at 0 K. Thus, the physical meaning of the Ewald summation method is that it determines the electrostatic potential and energy for an array of point charges at $T_{Gau} > 10000$ K for commonly used values [6,7] of the width of the Gaussian.

We can confirm the physical meaning of the Ewald sum method as follows. For this we consider the case when the width of the Gaussian, η, is small or the equivalent temperature, $T_{Gau}$, is sufficiently low that the Gaussians on different point charges do not overlap. The Ewald sum method <u>always gives the electrostatic potential and energy for an array of point charges</u> (and not charge distributions) for all values of the width of the Gaussian, η. But for the case considered above, the mathematical contribution from the second step vanishes and the Ewald method reduces to the first step (Fourier method), which gives the electrostatic potential and energy for an array of non-overlapping Gaussians centered on point charges. For example, the potential of the first step of the Ewald method, in Ref.3 and in Eq.4.24 of Ref.7 is identical to Eq.14 above. However, as seen earlier, Gausssian smearing of point charges implies that the crystal is at finite temerpature, $T_{Gau}$. Therefore, when η is small, the Ewald method, since it becomes equal to the Fourier method, gives the electrostatic potential and energy for an array of point charges at $T_{Gau}$. Clearly, the Ewald method cannot have two different physical meanings, a) the electrostatic potential and energy for an array of point charges at $T_{Gau}$ when the width of the Gaussian, η, is small and b) the electrostatic potential and energy for an array of point charges at 0 K when the width of the Gaussian, η, is large. Hence, the special case of small width of the Gaussian, η, confirms the physical meaning of the Ewald sum method, i.e. it gives the electrostatic potential and energy for an array of point charges at $T_{Gau}$. Therefore, in the use of the Ewald method, in addition to the width of the Gaussian that is always mentioned, the equivalent temperature, $T_{Gau}$, must also be mentioned.

The relation between the Fourier and Ewald formalisms when Gaussian smearing is used is also clear. The Fourier method is the first step of the Ewald method. For small values of the width of the Gaussian, η, the first step of the Ewald method (which is the Fourier method) suffices to determine the electrostatic potential and energy of an array of point charges. Whenever the width of the Gaussian, η, becomes sufficiently large that Gaussians on different point charges overlap, the Fourier method is no longer sufficient. The Ewald method is a generalization of the Fourier method with Gaussian smearing for large widths such that the Gaussians centered on different point charges overlap. It allows the determination of the electrostatic potential and energy of an array of point charges for all values of the widths of the Gaussian, η, or equivalently at all temperatures $T_{Gau}$.

As seen for Al and KCl earlier and true in general (unless demonstrated otherwise) the Gaussians on different point charges do not overlap near $T_p$, Therefore, the potential due to the Gaussian distributions are equal to the potential due to point charges. The first step of the Ewald sum method gives the electrostatic potential due to point charges. There is no need at all for the second step of the Ewald sum method for $T_{Gau} \leq T_p$. Since the electrostatic potential can be obtained from the Fourier method, it follows (from the discussion in the previous section) that for all realistic values of the width of the Gaussian, Eq.19 can also be used to determine the electrostatic potential, especially since it converges faster than Eq.14 as seen from Table 1.

## 5. Conclusion

The electrostatic potential and energy of point charges in a real crystal, in the presence of thermal vibrations, is obtained as a special case of the Fourier method. Incorporating the role of thermal vibrations in electrostatic energy calculations leads to the physical meaning of the Ewald sum method. The Ewald summation method determines the electrostatic potential and energy of point charges in a crystal at a temperature that is obtained from the width of the Gaussian and not at 0 K. For values of the width of the Gaussian commonly recommended for computational convenience temperatures exceed 10000 K.

**Table 1** Madelung constant in KCl expressed as the potential at the site of the potassium ion, V(K). V(K)$^A$ and V(K)$^B$ represent the potential obtained from Eq.14 and Eq.19 respectively. The width of the Gaussian used in both series is $\eta^2 = (2\langle u_s^2 \rangle) = a^2/225$ or $\eta = a/15$ and corresponds to a temperature near the melting point of KCl.

| h k l | $\lvert G \rvert^2 * a^2$ | m | $F_G$ | $e^{-\pi^2(2\langle u_s^2\rangle)G^2}$ | V(K)$^A$ (e/a) | $e^{-2\pi^2(2\langle u_s^2\rangle)G^2}$ | V(K)$^B$ (e/a) |
|---|---|---|---|---|---|---|---|
| 111 | 3 | 8 | 8 | 0.8767 | -10.97238 | 0.7686 | -6.74903 |
| 311 | 11 | 24 | 8 | 0.61723 | -7.54308 | 0.38097 | -4.63236 |
| 331 | 19 | 24 | 8 | 0.43455 | -6.14529 | 0.18884 | -4.02495 |
| 511, 333 | 27 | 32 | 8 | 0.30594 | -5.22194 | 0.0936 | -3.74245 |
| 531 | 35 | 48 | 8 | 0.2154 | -4.46970 | 0.0464 | -3.58042 |
| 533 | 43 | 24 | 8 | 0.15165 | -4.25417 | 0.023 | -3.54774 |
| 551,711 | 51 | 48 | 8 | 0.10677 | -3.99828 | 0.0114 | -3.52042 |
| 553,731 | 59 | 72 | 8 | 0.07517 | -3.76469 | 0.00565 | -3.50286 |
| 733 | 67 | 24 | 8 | 0.05292 | -3.71642 | 0.0028 | -3.50030 |
| 555,751 | 75 | 56 | 8 | 0.03726 | -3.64558 | 0.00139 | -3.49766 |